\documentclass[gmd, manuscript]{copernicus}

\usepackage{booktabs}
\usepackage{hyperref}
\usepackage{tikz}
\begin{document}

\title{\textit{``What is a realistic forecast?''} \\
Assessing data-driven weather forecasts, \\
a journey from verification to falsification.
}

\Author[][zied.benbouallegue@ecmwf.int]{Zied}{Ben-Bouall\`egue} 
\affil[]{European Centre for Medium-Range Weather Forecasts, Reading, United Kingdom}

\runningtitle{What is a realistic forecast?}
\runningauthor{Z. Ben-Bouall\`egue }

\received{}
\pubdiscuss{} 
\revised{}
\accepted{}
\published{}

\firstpage{1}

\maketitle
\nolinenumbers

\begin{abstract}
The artificial intelligence revolution is fuelling a paradigm shift in weather forecasting: forecasts are generated with machine learning models trained on large datasets rather than with physics-based numerical models that solve partial differential equations. This new approach proved successful in improving forecast performance as measured with standard verification metrics such as the root mean squared error. At the same time, the realism of data-driven weather forecasts is often questioned and considered an Achilles' heel of machine learning models.
How \textit{forecast realism} can be defined and how this forecast attribute can be assessed are the two questions simultaneously addressed here.
Inspired by the seminal work of \cite{murphy1993} on the definition of \textit{forecast goodness}, we identify 3 types of realism: a functional realism measured by scoring functions, a structural realism related to the statistical characteristics of the forecasts, and a physical realism that is apprehended through the lenses of our scientific knowledge.
This conceptual setting serves as a basis for the design of a new framework for the evaluation of data-driven weather models where falsification arises as a complementary process to the well-established diagnostic and verification tasks.
\end{abstract}

\section{Introduction}
\label{intro}
In operational centers, weather forecasts have taken the form of computer simulations since the 1960s \citep{lynch2008}. The concepts underlying \textit{numerical weather prediction} were already in place in the early 1900s with the seminal work of \cite{bjerknes1904}. They promoted  a two-step approach to weather forecasting: first, a diagnostic step that assesses the atmospheric state at a time $t_0$, followed by a prognostic step that evolves the system to time $t_0+\Delta t$. \cite{bjerknes1904} suggested that the evolution of the system should follow the laws of physics, the so-called primitive equations. His suggestion was successful.\\

With increased computer resources, observation capabilities, and sustained modelisation efforts, weather forecasting has indeed been through the so called quiet revolution \citep{bauer2015}. During the past 30 years, the gain in forecast accuracy was around 1 day of lead time per decade. With the rise of machine learning (ML) as a powerful tool for prediction, data-driven models have also demonstrated their ability to generate skillful weather forecasts \citep{zbb2024lameteo}. Today, these new types of models are becoming operational \citep[as for example at the European Centre for Medium Range Weather Forecasting,][]{boucher2025lameteo} and incidentally question the need to use primitive equations for weather forecasting.\\

Theory-driven forecasts and data-driven  forecasts are both computer simulations, that is, computer-generated representations of a target\footnote{The \textit{target} is also called indifferently \textit{truth} or \textit{observation} in the following.}. In both cases, a two-step approach is followed, with specific rules that make an initial state evolve with time.  Yet, the change of paradigm coincides with a change of logic: with theory-driven models, the logic applied is mainly deductive (the rules are defined following physics theories), while with data-driven models, the core logic is inductive (the rules are learned from data).\\

During the \textit{training} of ML models, induction is indeed at play. A set of rules is learned from observations  to describe, broadly speaking, the relationship between a situation at time $t_0$ and a situation at time $t_0+\Delta t$. Deductive logic is then applied in \textit{inference} mode: a dynamical system evolves following prescribed rules. The new set of rules is machine-readable and is used as such in inference mode. These ML rules can be considered as the primary output of an ML model, but they are not directly human-readable, hence the need for assessing the validity of these new theories as part of the assessment of ML models.\\

\begin{figure*}
\begin{center}
\includegraphics[trim={0.1cm 0 0 0cm},clip,width = 0.435\textwidth]{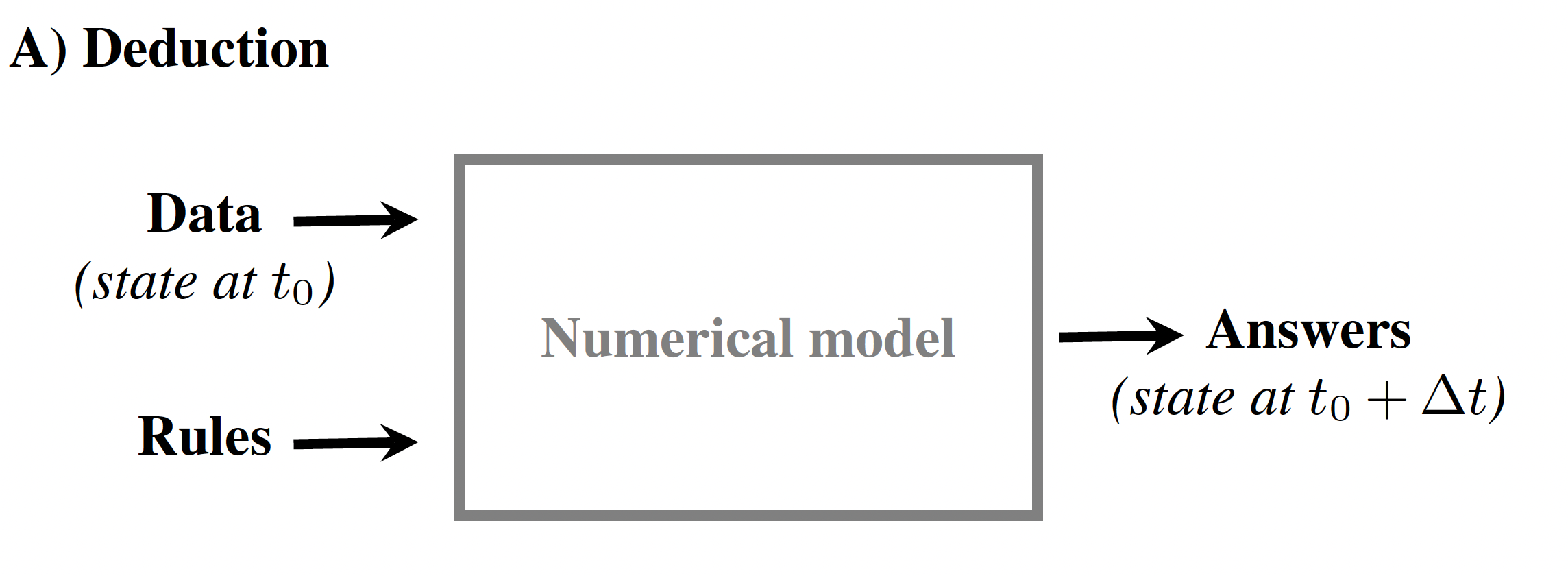}
\hspace{1.5cm}
\includegraphics[trim={0.1cm 0 0 0cm},clip,width = 0.40\textwidth]{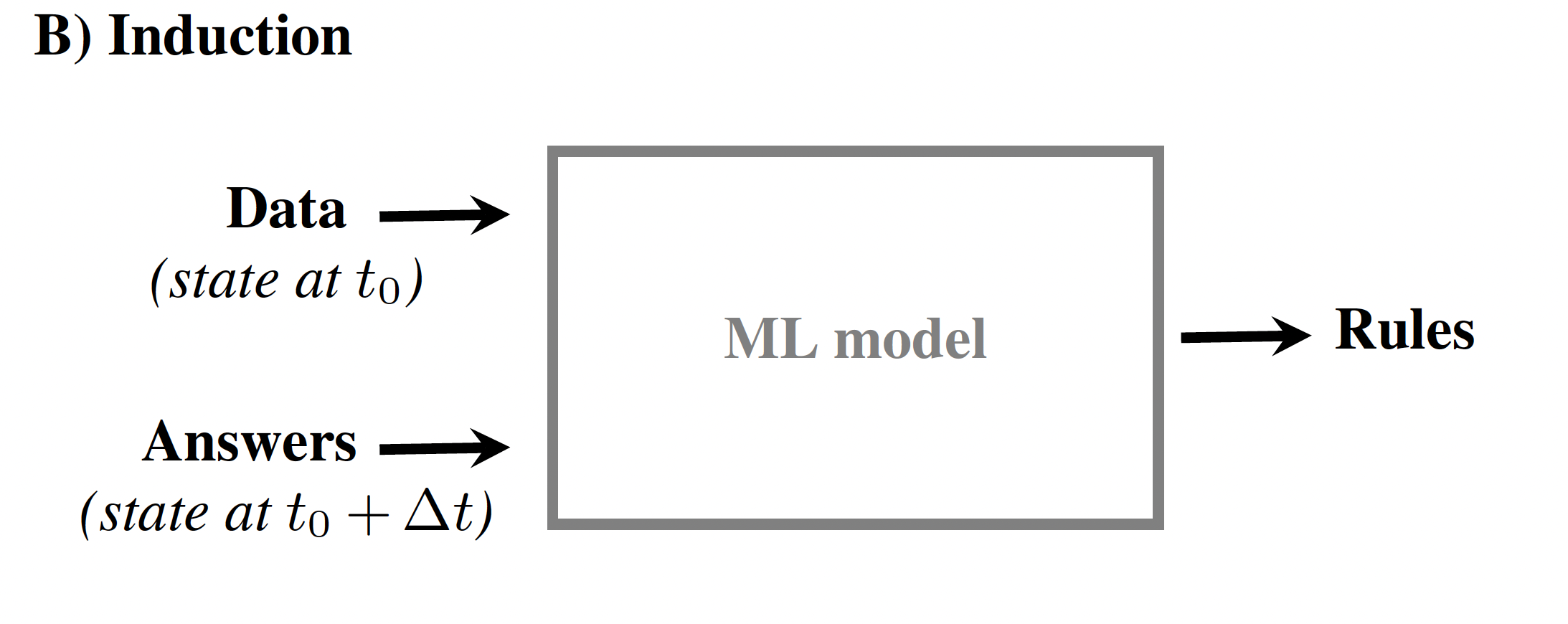}
\end{center}
\caption{Schematic of the two types of logic applied in weather forecasting: A) deduction and B) induction. In A), the rules of the numerical model are derived from the laws of physics for theory-driven models or as an output of B) for data-driven models.
}
\label{fig:logic}
\end{figure*}

The two types of logic discussed above are illustrated in Fig.~\ref{fig:logic}. In forecasting mode, deduction is applied using as input prescribed rules (laws of physics or the so-called \textit{weights} from ML models) and initial conditions (the data of  today) to provide an answer to the question ``What will be the weather tomorrow''. This is illustrated in Fig.~\ref{fig:logic}(A).  In training mode, induction is applied to build a set of rules that describes the relationship between the weather of today and the weather of tomorrow.  This is illustrated in Fig.~\ref{fig:logic}(B).\\

Interestingly, induction is also widely used when developing theory-driven models.  For example, parameter tuning and calibration are typically performed using this type of logic. On the other hand, data-driven models are not theory-free but rather epistemically charged through data collection and pre-processing as well as model design choices \citep{andrews2023}. The \textit{problem of induction}\footnote{First and foremost, an intriguing philosophical question formulated by \cite{hume1793}} becomes, in this context, a problem of interpretability with the challenge for a human to make sense of theories generated by ML. The other challenge, discussed here, is to make sense of the forecasts.\\

Indeed, with the advance of ML solutions and promising results in terms of predictive skill, forecast \textit{realism} has come out as a central point of discussion when assessing data-driven weather forecasts. Often, imperfect or limited realism is perceived as a potential blocker that can hinder the uptake of ML-based forecasts. However, in the literature, the term \textit{realism} can take on several meanings depending on the context and intent of the author. So, even if realism is acknowledged as a key attribute, no clear definition exists so far. This work aims at filling this gap.\\

The seminal work of \cite{murphy1993} on \textit{forecast goodness} was motivated in its time by the lack of clarity around the definition of a \textit{good} forecast.  He suggested 3 types of \textit{forecast goodness} related to 1) the forecast consistency with the forecaster's true belief, 2) the closeness of the forecast conditions to the observed conditions, and 3) the forecast value in a decision-making framework. This categorisation provides a way to organize one's thoughts when dealing with the complex task of assessing a weather forecast. Over the years, research activities have shed new light on the link between true belief and optimal score, as well as the link between forecast value and forecast quality. Nevertheless, Murphy's general verification framework remains a guiding reference when discussing \textit{forecast goodness} in an operational context \citep{harrison2025}.\\

Today, approaching the problem of verification with data-driven forecasts in mind, we try to answer the question ``What is a realistic forecast?''. For this, we suggest distinguishing 3 types of  \textit{forecast realism} that relate to 1) the forecast closeness to observations in each instance, 2) the average consistency of the forecast with the observation, and 3) the compatibility of the forecast with our physical knowledge. The first 2 types are commonly assessed as part of routine verification and diagnostic activities in place at operational centres. We argue that the third type, physical realism, becomes particularly relevant when dealing with data-driven forecasts. More generally, this is the case whenever statistical methods are involved, such as stochastic perturbations for ensemble generation or post-processing for bias correction.\\

{\color{black}Our contribution consists of relating 2 forms of realism to already well-established concepts in the forecast verification literature, namely the concepts of scoring function and calibration. In doing so, we highlight that improving forecast scores and improving calibration are generally aligned objectives in a probabilistic forecasting framework. By contrast, an accuracy–activity trade-off emerges when dealing with single deterministic forecasts.\\

Our discussion extends to a third form of realism apprehended through our scientific knowledge. We argue that providing clear definitions of these different forms of realism will facilitate the development of methods tailored to each. For example, forecast smoothness is a characteristic that contributes to a form of realism of the forecast, yet measures of smoothness provide little information about whether the forecast is physically consistent. Distinguishing between these different forms of realism helps not only to support appropriate verification practices but also to provide more actionable feedback to model developers. }\\

The manuscript is organised as follows. In Section \ref{sec:realism}, we introduce the concept of functional, structural, and physical realism. We discuss how these 3 types of realism can be measured or assessed, as well as their relationship. In Section \ref{sec:discuss}, we expand the discussion to touch on closely related topics such as forecast value and model validation, before closing the paper with the description of a typical \textit{data-driven forecast evaluation journey}.

\section{Types of forecast realism}
\label{sec:realism}

\subsection{Type 1 of realism and Forecast Verification}
\label{sec:type1}
The first type of realism is related to the notion of closeness to reality.
We use the term \textit{functional realism} to refer to the ability of a forecast to be close to the truth on a given occasion. Functional realism can be assessed with a scoring function that measures a distance between a forecast and an observation.\\

Let's denote $x_i$ a forecast and $y_i$ the corresponding observation for an instance $i$. A verification measure $V$ would follow:
\begin{equation}
   V= v(x_i,y_i),
 \label{eq:verif}
\end{equation}
with $v$ a scoring function.  The forecast $x$ can be a direct model output or a transformation of it (\textit{e.g.} hourly or daily precipitation), a scalar or a vector (\textit{e.g.} wind speed or wind components), a point forecast or an ensemble, in real value space or probability space, and so on. Measuring $V$ is part of a process described as \textit{forecast verification}.\\

The score $V$ is estimated for each instance $i$ of the forecast.  When $V$ is averaged over a verification sample, it is often referred to as a measure of \textit{forecast accuracy}.  Accuracy differs from forecast quality as the latter encompasses more aspects such as \textit{forecast association} or \textit{forecast bias} that are not direct measures of a distance between forecast and observations.\\

Key concepts related to the realism of Type 1 are the concepts of elicitation and scoring rules. While a formal description of these mathematical tools is beyond the scope of this paper, we can stress their implication on how verification is performed today. Indeed, common practice involves assessing an \textit{elicitable functional} with a \textit{consistent scoring function},  such as, for example,  computing the squared error of an ensemble mean. 
Generally speaking, proper scores applied to specific functionals offer a theoretical guarantee of an alignment of meaning (calibration) and scope (optimal score).\\

Ultimately, the way one approaches functional realism should reflect the application one has in mind. In  weather forecast verification, each result reflects not only the choice of metric but also the choice of substrate (\textit{e.g.} variable, domain, lead time, spatial scale).  A primary purpose of forecast verification is to quantify and communicate forecast skill, that is to what extent a forecast is better than a reference, and to compare and rank competing models. When combined with additional information, such as known differences between competing models, or when expressed through conditional scores, verification results can provide valuable diagnostic insights too.

\subsection{Type 2 of realism and Model Diagnostics}
\label{sec:type2}
The second type of realism is related to the idea of statistical consistency between forecast and observation. On average, one expects to see a forecast with the same statistical characteristics as the observation.\\

Let's denote $X$ and  $Y$ the characteristics of a forecast and of the corresponding observation, respectively. A diagnostic measure $D$ would follow:
\begin{equation}
   D = d(X,Y),
\label{eq:diag}
\end{equation}
with $d$ a summary measure of the difference between $X$ and $Y$. For example, if $X$ and $Y$ represent the mean forecast and the mean observation, respectively, and $d$ is a simple subtraction function, then $D$ is called \textit{bias}. The same concept applies to the variance (or variance of the anomalies) instead of the mean, and the result is then referred to as \textit{activity bias}.\\

The ability of a forecast to be statistically consistent with the observations is often referred to as \textit{forecast reliability}. A reliable forecast would be statistically indistinguishable, or rather drawn from the same underlying probability distribution as the observation. Reliability is a key forecast attribute: when a forecast is reliable, a user can take it at face value. In other words, there is no need for a mental (or statistical) adjustment before using it for decision-making.\\

Forecast bias and forecast activity bias are simple measures of forecast reliability. Numerous other metrics and tools exist. For example, power spectra are  popular for comparing the energy level at various scales in deterministic or ensemble forecasts \citep{rodwell2025}. With this approach, one can assess the effective forecast resolution, that is, the smallest spatial scale at which atmospheric structures are reproduced with realistic amplitudes \citep{selz2025}.
Also, as the first generation of data-driven forecasts were smoother
at longer lead times, a set of methods were developed to measure and compare characteristics such as granularity and sharpness of spatial fields \citep{ebert2025}.\\

The term \textit{model diagnostics} is used here to describe the process of assessing forecast reliability. This process helps identify model weaknesses, \textit{e.g.}, regional biases, too low forecast activity, or not enough ensemble spread. A given model, rather than a specific forecast, is at the heart of diagnostic investigations. The diagnostic results inform developers on how to improve their forecasting model. In a traditional sense,
\textit{diagnostic} activities go beyond a forecast reliability assessment to include the variety of methods deployed to understand the origin of forecast errors \citep{magnusson2017}.

\subsection{Relationship between Type 1 and Type 2}
\label{sec:type1v2}

Does an improvement in Type 2 of realism  automatically mean an improvement in Type 1? In other words, does improving the reliability of a forecast lead to better forecast accuracy as measured by a scoring function? The link between functional and structural realism has different implications in a probabilistic and in a deterministic framework. \\

A proper scoring rule allows for a decomposition into terms related to the resolution and the reliability of a forecast \citep{broecker2009}. In a probabilistic framework, improving Type 2 goes hand-in-hand with improving Type 1 when using proper scoring rules: improving reliability leads to better scores. In passing, it is probably for this reason that \textit{propriety} is often considered a fundamental score property.\\

Things are different in a deterministic framework. 
A first moment calibration (bias correction) of the forecast leads to an improvement of the score, while a second moment calibration (variance correction) can lead to its degradation. This well-known result is related to the predictability of the system, as demonstrated in \cite{murphy1988}. As a consequence, improving Type 1 and improving Type 2 can appear as contradictory goals, a conundrum referred to as the \textit{accuracy versus activity} trade-off in \cite{zbb2024blog}.\\

{\color{black}
Data-driven models trained to minimise the mean squared error produce forecasts that are indeed smoother than their physics-based counterparts. During the optimisation process, small-scale features with limited predictability are effectively averaged out, resulting in a loss of variability at fine scales. The level of smoothness depends on the roll-out strategy: the longer the training roll-out, the lower the predictability, and the smoother the forecast. As a result of the optimisation process, a data-driven forecast is equivalent, to some extent, to a conditional mean up to a certain forecast range. \\

The outputs of physics-based weather models are interpreted as plausible trajectories given the physical laws and the chaotic nature of the Earth's system.
In contrast, the outputs of data-driven models are solutions to a statistical optimisation problem defined by a training objective. For this reason, the former is called a ``deterministic'' forecast while the latter is a new type of information support often referred to as a ``single'' forecast with ideally a fine balance between realism of type 1 and 2.  \\}

\subsection{Type 3 of realism and Theory Falsification}
\label{sec:type3}

Type 3 of realism or \textit{physical realism} is related to the ability of a forecast to be plausible in regard to our understanding of the laws of Nature.
The laws of physics set a clear demarcation between outcomes that are possible and outcomes that are not. A forecast is physically realistic when it belongs to the former category.  \\

Let's denote $\mathcal{K}$ our knowledge base. This knowledge includes the laws of physics that govern the weather of our planet and, more generally, the physical properties of the Earth system. A so called \textit{falsification} test $F$ applied to a forecast $x_i$ for an instance $i$ would follow:
\begin{equation}
   F = f(x_i,\mathcal{K}), 
\label{eq:fals}
\end{equation}
with $f$ a hypothesis test function which assesses if a forecast $x_i$ is compatible with scientific knowledge $\mathcal{K}$.\\

We refer to the process of checking for physical realism as \textit{falsification}. This terminology is borrowed from the field of philosophy of science\footnote{Falsification is a concept first introduced in \cite{popper}.} where falsification is originally applied to scientific theories. With the use of observations, a rejectionist approach is followed: a scientific theory is falsified if an observation is outside the limits drawn by this theory. Here, the process is somehow reversed as we start from the observations: a forecast is falsified if it falls outside what is deemed possible according to consolidated physics theories.  \\

{\color{black}
Forecast falsification applies to direct model output as well as to derived quantities and compound forecasts. The forecast is checked for physical realism in each instance rather than as in terms of statistical expectations. For example, when snowfall is forecast, near-surface temperatures are expected to be close to freezing. Demonstrating a strong statistical relationship between snowfall and temperature is not, by itself, evidence of physical realism; rather, physical realism requires consistency in all instances. Falsification is also applied to check the generalization capabilities of ML models, that is, their ability to remain physically consistent in situations different from the training sample or in controlled experiments.  \\

In addition, verification practitioners are facing a new class of errors associated with generative ML models: \textit{hallucinations}. These errors correspond to forecast features that appear statistically plausible but are clearly inconsistent with the underlying physical processes when examined by domain experts \citep{rathkopf2025}. Because conventional scoring rules are insensitive to physical realism, such hallucinations usually go unnoticed with common verification practices. Detecting these errors, therefore, requires new tools. As a bigger challenge, a formal relationship between improvements in physical realism and improvements in forecast accuracy, as measured by standard scoring functions, remains to be established.\\

Here, we advocate developing testing protocols specifically designed to assess physical realism. Such protocols would provide two main benefits. First, a quantitative assessment of physical realism would enable the comparison of forecasting systems and help define acceptable levels of realism for specific applications. Second, falsification or invalidation of the model is a crucial step in the scientific process of model development \citep{Beven2019}. Identifying forecasts that violate fundamental physical constraints can provide model developers with actionable information to guide future improvements, such as implementing physical constraints in the model.
\\
}

Interestingly, \textit{Type 3 of realism} is closely related to \textit{Type 1 of goodness}. In a bold statement, \cite{murphy1993} stresses that ``a forecast should always correspond to a forecaster's best judgment'', where  ``a forecaster's judgment on a particular occasion is assumed to contain \textit{all} of the information in her knowledge base on this occasion''. Here, we reformulate this principle considering the knowledge base in general rather than in a specific situation: a forecast should always align with a forecaster's scientific knowledge. Cases where this condition is not met should be reported to both model developers and forecast users.

{\color{black}
\subsection{Falsification studies}
\label{sec:type3}

Forecast falsification can take several forms, but first and foremost, it encompasses the inspection of direct model outputs. Visualisation of spatial fields, vertical profiles, time series, or forecast animations  can be complemented with the analysis of case studies focusing on specific events as discussed in Section \ref{sec:case}. The examination of forecasts by knowledgeable experts has a long tradition in meteorology and is often referred to as subjective verification \citep{stanski1989}. \\

Beyond visual inspection, we distinguish three additional levels of investigation. The first additional level consists of checking that the forecast respects physical consistency relationships and that no forecast is found in ``no-go areas'' defined by fundamental physical constraints.
No-go areas exist for individual variables (\textit{e.g.}, negative precipitation) or for combinations of variables (\textit{e.g.}, positive snowfall when total precipitation is zero). In an operational setting, monitoring these hard constraints could be integrated into existing verification workflows by archiving forecast extrema alongside standard verification metrics. \\

The second level of investigation is related to the physical concept of equilibrium, from which are derived conservation laws, dynamical and thermodynamical balances.
Unlike simple physical constraint checks, assessing a level of equilibrium can be done quantitatively, enabling the definition of performance metrics \citep{kasteleyn2026}.  A practical limitation of this approach is that ML models output only a restricted set of variables. As a result, some conservation laws and balance relationships cannot be evaluated directly, limiting the range of physical diagnostics that can be applied.\\

The third level of investigation requires access to the forecasting model itself in order to conduct targeted tests, such as dynamical and generalisation tests. These tests involve running the model under perturbed or controlled conditions designed to elicit specific responses, and evaluating the model's ability to respond adequately. For example, ML models run in inference for longer lead times than their typical forecast range can reveal unexpected driftss as in \cite{lehmann2026}. In \cite{hakim2024}, the encoding of realistic physics was assessed by analysing the impact of local perturbations on the resulting dynamical flow. Such falsification studies are typically conducted by model developers and can be approached as stress tests that are essential for building trust in their forecasting systems \citep{hahner2026}.
}

\section{Discussion}
\label{sec:discuss}
\subsection{Perfect forecasts}
\label{sec:perfect}
Computer simulations are idealizations, so they are incorrect representations of the truth to some extent.
Similarly, observations are representations of the truth with their own imperfections.
However, a thought experiment set in a perfect world can be useful for highlighting the hierarchical relationship among the 3 types of realism.\\

First, let's consider a perfect Type 1 forecast. In that case, forecasts and observations are identical in all instances, which also leads to perfect realism of Types 2 and 3. Now consider a perfect realism of Type 2 such as forecasts and observations are drawn from the same distribution. In that case, forecasts and observations are not necessarily identical (no Type 1 perfection), but a forecast is always taking a value that could be observed in principle. In other words, a forecast can be regarded as a counterfactual of the observation. Finally, consider a perfect realism of Type 3. Following the laws of physics does not guarantee either perfect reliability (as observed in theory-driven forecasts) or a perfect match with the observations.

\subsection{Fit-for-purpose?}
The hierarchy derived from the perfect forecast thought experiment in Section \ref{sec:perfect} does not necessarily reflect the importance of the 3 types of realism for a given user or a given application. The guiding question ``Is a forecast/model fit for purpose?'' can help set priorities.\\

The relative importance of the different forms of realism ultimately depends on the intended application. For example, Type 1 realism may be particularly important for day-to-day forecasting applications, where the priority is to produce forecasts that closely match observations in specific situations and at short lead times. Type 2 realism may become more important at longer time scales, where forecasts are expected to reproduce the statistical and structural characteristics of the observed system. Calibration is also a necessary condition for optimal decision-making based on probabilistic forecasts.\\

{\color{black} Type 3 of realism is relevant to scientific exploration when one expects fundamental laws of physics to be respected.  Another example is climate projection. In a changing climate, consistency with physical knowledge is considered a key attribute for effective generalisation beyond the conditions seen in the training sample. More generally, forecasts that violate physical laws can undermine confidence in the forecasting system. This aspect is key in applications with a ``human in the loop’’, where trust plays a central role. If the forecast serves as input to downstream post-processing systems, inconsistencies can be corrected or mitigated at this later stage.\\
}

\subsection{Validation and case-study analysis}
\label{sec:case}
Checking whether a model is fit-for-purpose is often referred to as the \textit{validation} step. As part of the evaluation of computer simulations, \textit{validation} assesses the practical implementation of the ideas used to build a model \citep{winsberg2010}. In ML, the validation step consists of checking the statistical performance of a trained model on an unseen dataset, \textit{e.g.}, with no overlap between training and validation periods. \\

A validation exercise can take the form of an in-depth analysis of a case study. Strengths and weaknesses are noted while assessing all three types of realism.  For example, storm Ciaran was discussed in \cite{charlton2024} revealing that the track of the storm was well captured by the different models (Type 1), but the maximum wind in the ML forecasts was generally underestimated compared to observations (Type 2), and the physical processes at play during the strengthening of the storm were sometimes poorly represented (Type 3).

\subsection{Trust and interpretability}
Arguably, physical realism is sometimes more a desired property than actually needed in practice. Artifacts exist in the output of both theory-driven and data-driven forecasts, but Type 3 of realism is key in building trust when it comes to ML forecasts as a consequence of the \textit{induction problem} mentioned in Section \ref{intro}. To complement falsification studies that focus on the output of ML models, \textit{interpretability} studies investigate ML models themselves. Making ML models \textit{interpretable} is an area of research that aims to open the ``black box'' \citep{mcgovern2019,molnar2025}.   \\

Ideally, the new forecasting rules derived with ML models should be interpretable by humans, not just by machines. Simultaneously, there is a growing interest in \textit{explainability} which refers to ``the extent to which humans can comprehend and rationalise the predictions of an ML model'' \citep{laloyaux2025}.  Nowadays, ethical considerations also encompass discussions on robustness, reproducibility, and fairness \citep[see, for example,][]{olivetti2025}.

\subsection{Realism and information content}
Type 1 of realism and \textit{information content} are two forecast attributes that are closely related. A formal link  between them can be demonstrated with the help of post-processing.  If we first calibrate a forecast and then assess its performance with scoring rules, we are effectively measuring its \textit{resolution} or information content. This approach is recommended for a fairer comparison of forecasts with different reliability properties, leading to a comparison of their potential performance \citep{gneiting2025}. In this context, post-processing becomes a key component of the verification process. \\

Assessing the information content of a forecast is a step towards assessing its value \citep[Type 3 of \textit{goodness} in ][]{murphy1993}. The usefulness of a forecast in a decision-making process can be measured using cost-loss models, for example \citep{richardson2000}. Another aspect of the forecast value revolves around the idea of complementarity.
When forecasts from different origins are available, what matters is if their information content is complementary rather than if their performance are similar as measured with a given metric. A forecast with less realism can be of greater value in this context.

\subsection{Knowledge base and progress}

What is the place of \textit{scientific knowledge} in weather forecasting? How can a better understanding of the Earth system lead to better forecasts while using ML as a core technology? Here, we argued that scientific knowledge should have a special place in the process of evaluating ML forecasts, through falsification. \\

{\color{black} Falsification tools, applied to the outputs of data-driven models, complement traditional verification methods by assessing aspects of forecast realism that are not captured by statistical scores alone. Beyond their value for evaluation, such tools can also support model development by identifying physically implausible behaviours and guiding the design of physically informed model architectures. In this way, falsification serves not only as a framework for assessing physical realism but also as a pathway toward more trustworthy and scientifically grounded forecasting systems. \\ }

Physical understanding of the Earth system already guides model developments. For example, physical constraints can enter the design of the loss function or the model architecture \citep{sha2025}. As illustrated in \cite{moldovan2025}, negative precipitation can be prevented by implementing appropriate architecture choices. This approach contrasts with the use of a physics-based post-processing step, which consists of setting negative forecast values to zero before dissemination.  \\

Data are also epistemically charged. From the infancy of the Scientific Method up to today, knowledge and data 
 are not independent but rather continuously feed and define one another. In weather forecasting, the output of a numerical model based on physical laws  (\textit{Answers} in Fig.~\ref{fig:logic}A) can serve as input of a machine learning model  (\textit{Data} and \textit{Answers} in Fig.~\ref{fig:logic}B). As a typical example, global weather models are trained using ERA5 \citep[the fifth generation ECMWF atmospheric reanalysis, produced by the Copernicus Climate Change Service,][]{era5}.

\begin{figure}[h]
\vspace{0.2cm}
\begin{center}
\includegraphics[trim={0cm 0 0 0cm},clip,width = 0.33\textwidth]{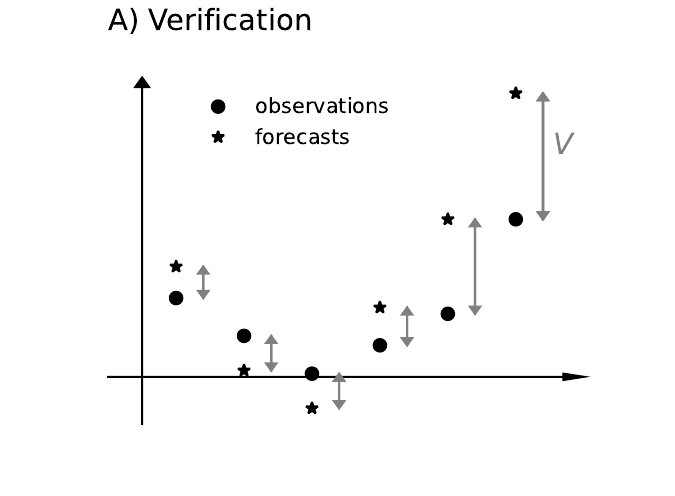}
\includegraphics[trim={0cm 0 0 0cm},clip,width = 0.33\textwidth]{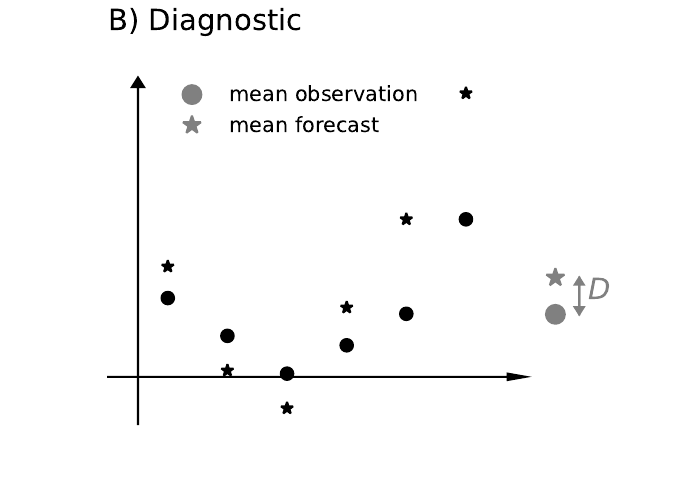}
\includegraphics[trim={0cm 0 0 0cm},clip,width = 0.33\textwidth]{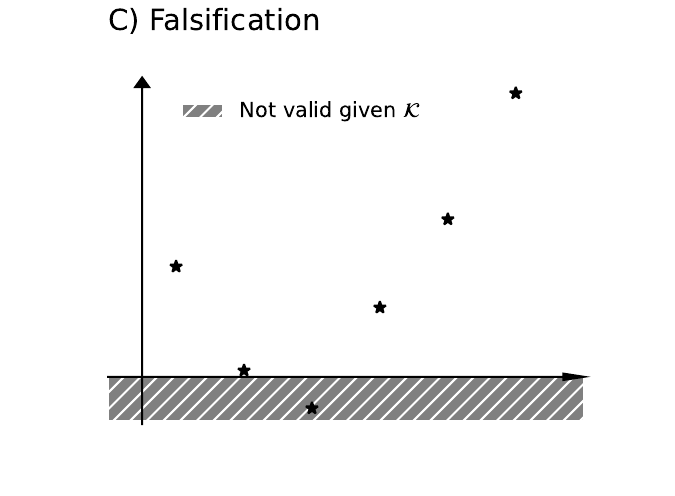}

\caption{Illustration of evaluation activities to assess the 3 types of realism discussed in this manuscript: A) verification $V$ to assess the functional realism, B) diagnostic $D$ to assess the structural realism, and C) falsification based on knowledge base $\mathcal{K}$ to check for physical realism.  }
\label{fig:journey}
\end{center}
\vspace{-0.4cm}
\end{figure}

\subsection{A typical journey}
\label{sec:conc}

{\color{black} Take the example of a bivariate forecast. Functional realism can be measured using scoring rules applied to each variable separately, but also computing multivariate scores in the bivariate space. Structural realism, in contrast, requires evaluating whether the forecast reproduces the statistical properties of the observations, including the marginal distributions of individual variables, their tail behaviour, and the dependence structure between variables. Physical realism adds a further layer of assessment by examining whether the forecasts remain within physically plausible bounds and whether fundamental physical constraints between the 2 variables are respected.\\ }

Fig. \ref{fig:journey} schematically illustrates  the 3 steps we have proposed for evaluating data-driven weather forecasts. So, a typical journey would go through the following stations:
\begin{itemize}
    \item  Verification: a comparison of a forecast with our perception of reality (observations) in each instance to help rank models and inform decisions in model development.
\item  Diagnostic:  an assessment of the average forecast characteristics to shape our understanding of the model deficiencies.
\item Falsification: a comparison of a forecast with our understanding of reality (scientific knowledge) to identify artifacts and build trust.
\end{itemize}
At the end of the line, one should have factual elements to answer the question ``Is the forecast realistic?''.\\

\bibliographystyle{ametsocV6}
\bibliography{references}

@misc{sha2025,
      title={Improving {AI} weather prediction models using global mass and energy conservation schemes},
      author={Yingkai Sha and John S. Schreck and William Chapman and David John Gagne II},
      year={2025},
      eprint={2501.05648},
      archivePrefix={arXiv},
      publisher={arXiv},
      primaryClass={physics.ao-ph},
}

@Inbook{Beven2019,
author="Beven Keith and Lane Stuart",
title="Invalidation of Models and Fitness-for-Purpose: A Rejectionist Approach",
bookTitle="Computer Simulation Validation: Fundamental Concepts, Methodological Frameworks, and Philosophical Perspectives",
year="2019",
publisher="Springer International Publishing",
address="Cham",
pages="145--171",
isbn="978-3-319-70766-2",
doi={10.1007/978-3-319-70766-2_6}
}

@article {murphy1993,
      author = "Allan H.  Murphy",
      title = {{What Is a Good Forecast? An Essay on the Nature of Goodness in Weather Forecasting}},
      journal = "Weather and Forecasting",
      year = "1993",
      publisher = "American Meteorological Society",
      address = "Boston MA, USA",
      volume = "8",
      number = "2",
      doi = "10.1175/1520-0434(1993)008<0281:WIAGFA>2.0.CO;2",
      pages=      "281 - 293",
      }

@article {murphy1988,
      author = "Allan H.  Murphy",
      title = "Skill Scores Based on the Mean Square Error and Their Relationships to the Correlation Coefficient",
      journal = "Monthly Weather Review",
      year = "1988",
      publisher = "American Meteorological Society",
      address = "Boston MA, USA",
      volume = "116",
      number = "12",
      doi = "10.1175/1520-0493(1988)116<2417:SSBOTM>2.0.CO;2",
      pages=      "2417 - 2424",
      url = "https://journals.ametsoc.org/view/journals/mwre/116/12/1520-0493_1988_116_2417_ssbotm_2_0_co_2.xml"
}

@ARTICLE{bauer2015,
author="P. Bauer and A. Thorpe and G. Brunet",
title=" The quiet revolution of numerical weather prediction",
journal="Nature",
year=2015,
volume=525,
pages={47--55}
}

@article{lynch2008,
title = {The origins of computer weather prediction and climate modeling},
journal = {Journal of Computational Physics},
volume = {227},
number = {7},
pages = {3431-3444},
year = {2008},
note = {Predicting weather, climate and extreme events},
issn = {0021-9991},
doi = {https://doi.org/10.1016/j.jcp.2007.02.034},
author = {Peter Lynch},
}

@article{charlton2024,
    author = "Charlton-Perez, A.J. and Dacre, H.F. and Driscoll, S. et al. ",
    title = {{Do AI models produce better weather forecasts than physics-based models? A quantitative evaluation case study of Storm Ciarán.}},
    journal = "npj Clim Atmos Sci 7, 93",
    year = 2024,
    doi="10.1038/s41612-024-00638-w"
}

@article {ebert2025,
      author = "Imme Ebert-Uphoff and Lander Ver Hoef and John S. Schreck and Jason Stock and Maria J. Molina and Amy McGovern and Michael Yu and Bill Petzke and Kyle Hilburn and David M. Hall and David John Gagne and William F. Campbell and Jacob T. Radford and Jebb Q. Stewart and Sam Scheuerman",
      title = {{Measuring Sharpness of AI-Generated Meteorological Imagery}},
      journal = "Artificial Intelligence for the Earth Systems",
      year = "2025",
      publisher = "American Meteorological Society",
      address = "Boston MA, USA",
      volume = "4",
      number = "3",
      doi = "10.1175/AIES-D-24-0083.1",
      pages=      "e240083",
}

@misc{stanski1989,
author = {Stanski, H.R. and L.J. Wilson and W.R. Burrows},
year=1989,
title={{Survey of Common Verification Methods
in Meteorology}},
publisher ={WMO Research Report No. 89-5}
}

@Article{bjerknes1904,
author = "Bjerknes, Vilhelm",
journal = "Meteorologische Zeitschrift",
year = 1904,
title = {{Das Problem der Wettervorhersage, betrachtet vom Standpunkte der Mechanik und der Physik
(The problem of weather prediction, considered from the viewpoints of mechanics and physics)}},
volume = "21",
pages = {1--7
(translated and edited by VOLKEN E. and S. BR ¨ONNIMANN. – Meteorol. Z. 18 (2009), 663–667)},
doi = "10.1127/0941-2948/2009/416",
}

@Article{zbb2024lameteo,
title={Prévisions météorologiques reposant sur l'intelligence artificielle : {Une} révolution peut en cacher une autre},
author={Zied {Ben Bouallègue} and Mariana Clare and Matthieu Chevallier},
journal = "La Météorologie",
pages={48--52},
doi= {10.37053/lameteorologie-2024-0058},
year=2024,
volume=126
}

@Article{boucher2025lameteo,
title={La prévision du temps par {IA} devient opérationnelle au {CEPMMT}},
author={Eulalie Boucher and Gabriel Moldovan and Zied {Ben Bouallègue} and Nina Raoult and Michael Maier-Gerber and Mariana Clare and Joffrey Dumont Le Brazidec },
journal = "La Météorologie",
pages={4--7},
doi= {10.37053/lameteorologie-2025-0051},
year=2025,
volume=130
}

@article{rodwell2025,
author = {Rodwell, Mark J. and Clare, Mariana C. A. and Lock, Sarah-Jane and Lonitz, Katrin and Chevallier, Matthieu},
title = {Power Spectra of Physics-Based and Data-Driven Ensembles},
journal = {Meteorological Applications},
volume = {32},
number = {5},
pages = {e70071},
doi = {https://doi.org/10.1002/met.70071},
year = {2025}
}

@article{selz2025,
title={On the effective resolution of {AI} weather prediction models},
DOI={10.22541/essoar.174139239.94807670/v1},
journal={ESS Open Archive},
author={Selz, Tobias and Bruinsma, Wessel and Craig, George C. and Markou, Stratis and Turner, Richard and Vaughan, Anna},
year={2025},
 }

@misc{zbb2024blog,
title={Accuracy versus activity},
DOI={0.21957/8b50609a0f},
publisher={AIFS Blog},
author={{Ben Bouallègue}, Zied and {the AIFS team}},
year={2024},
}

@misc{gneiting2025,
      title={{Probabilistic measures afford fair comparisons of AIWP and NWP model output}},
      author={Tilmann Gneiting and Tobias Biegert and Kristof Kraus and Eva-Maria Walz and Alexander I. Jordan and Sebastian Lerch},
      year={2025},
      archivePrefix={arXiv},
      primaryClass={stat.AP},
      url={https://arxiv.org/abs/2506.03744},
}

@book{hume1793,
  author = {David Hume},
  title = {A Treatise of Human Nature},
  year = {1793},
}

@book{popper,
  author = {Karl Popper},
  title= {{The Logic of Scientific Discovery}},
  year= 1959,
}

@book{winsberg2010,
  author = {Winsberg, Eric},
  title = {Science in the Age of Computer Simulation},
  year = {2010},
  publisher = {University of Chicago Press},
  address = {Chicago, IL}
}

@misc{laloyaux2025,
      title={Using data assimilation tools to dissect GraphDOP},
      author={Patrick Laloyaux and Mihai Alexe and Eulalie Boucher and Peter Lean and Ewan Pinnington and Simon Lang and Tobias Necker and Anthony McNally},
      year={2025},
      eprint={2510.27388},
      archivePrefix={arXiv},
      publisher={arXiv},
      url={https://arxiv.org/abs/2510.27388},
}

@article {mcgovern2019,
      author = "Amy McGovern and Ryan Lagerquist and David John Gagne and G. Eli Jergensen and Kimberly L. Elmore and Cameron R. Homeyer and Travis Smith",
      title = "Making the Black Box More Transparent: Understanding the Physical Implications of Machine Learning",
      journal = "Bulletin of the American Meteorological Society",
      year = "2019",
      publisher = "American Meteorological Society",
      volume = "100",
      number = "11",
      doi = "10.1175/BAMS-D-18-0195.1",
      pages=      "2175 - 2199",
}

@Article{moldovan2025,
AUTHOR = {Moldovan, G. and Pinnington, E. and Prieto Nemesio, A. and Lang, S. and Ben Bouall\`egue, Z. and Dramsch, J. and Alexe, M. and Santa Cruz, M. and Hahner, S. and Cook, H. and Theissen, H. and Clare, M. and O'Brien, C. and Polster, J. and Magnusson, L. and Mertes, G. and Pinault, F. and Raoult, B. and de Rosnay, P. and Forbes, R. and Chantry, M.},
TITLE = {{AIFS 1.1.0: An update to ECMWF's machine-learned weather forecast model AIFS}},
JOURNAL = {EGUsphere},
VOLUME = {2025},
YEAR = {2025},
PAGES = {1--23},
DOI = {10.5194/egusphere-2025-4716}
}

@article{broecker2009,
author = {Bröcker, Jochen},
title = {Reliability, sufficiency, and the decomposition of proper scores},
journal = {Quarterly Journal of the Royal Meteorological Society},
volume = {135},
number = {643},
pages = {1512-1519},
doi = {https://doi.org/10.1002/qj.456},
year = {2009}
}

@book{molnar2025,
  title={Interpretable Machine Learning},
  subtitle={A Guide for Making Black Box Models Explainable},
  author={Christoph Molnar},
  year={2025},
  edition={3},
  isbn={978-3-911578-03-5},
  url={https://christophm.github.io/interpretable-ml-book}
}

@article {hakim2024,
      author = "Gregory J. Hakim and Sanjit Masanam",
      title = "Dynamical Tests of a Deep Learning Weather Prediction Model",
      journal = "Artificial Intelligence for the Earth Systems",
      year = "2024",
      publisher = "American Meteorological Society",
      address = "Boston MA, USA",
      volume = "3",
      number = "3",
      doi = "10.1175/AIES-D-23-0090.1",
      pages= "e230090",
}

@article {harrison2025,
      author = "David R. Harrison and Amy McGovern and Christopher D. Karstens and Ann Bostrom and Julie L. Demuth and Israel L. Jirak and Patrick T. Marsh",
      title = "An Assessment of How Domain Experts Evaluate Machine Learning in Operational Meteorology",
      journal = "Weather and Forecasting",
      year = "2025",
      publisher = "American Meteorological Society",
      address = "Boston MA, USA",
      volume = "40",
      number = "3",
      doi = "10.1175/WAF-D-24-0144.1",
      pages=      "393 - 410",
}

@misc{andrews2023,
          author = {Mel Andrews},
           title = {The Devil in the Data: Machine Learning \& the Theory-Free Ideal},
           month = {October},
            year = {2023},
             url = {https://philsci-archive.pitt.edu/26075/}
}

@misc{rathkopf2025,
      title={Hallucination, reliability, and the role of generative AI in science},
      author={Charles Rathkopf},
      year={2025},
      eprint={2504.08526},
      archivePrefix={arXiv},
      publisher={arXiv},
      url={https://arxiv.org/abs/2504.08526},
}

@article{magnusson2017,
author = {Magnusson, L.},
title = {Diagnostic methods for understanding the origin of forecast errors},
journal = {Quarterly Journal of the Royal Meteorological Society},
volume = {143},
number = {706},
pages = {2129-2142},
doi = {https://doi.org/10.1002/qj.3072},
year = {2017}
}

@article{richardson2000,
author = {Richardson, D. S.},
title = {Skill and relative economic value of the ECMWF ensemble prediction system},
journal = {Quarterly Journal of the Royal Meteorological Society},
volume = {126},
number = {563},
pages = {649-667},
doi = {https://doi.org/10.1002/qj.49712656313},
year = {2000}
}

@article{olivetti2025,
title={Whose weather is it? A fairness framework for data-driven weather forecasting},
DOI={10.1088/1748-9326/ae21f5},
journal ={Environmental Research Letters},
volume={20},
number={12},
author={Olivetti, Leonardo and Messori, Gabriele},
year={2025},
}

@article{era5,
author = {Hersbach, Hans and Bell, Bill and Berrisford, Paul and Hirahara, Shoji and Horányi, András and Muñoz-Sabater, Joaquín and Nicolas, Julien and Peubey, Carole and Radu, Raluca and Schepers, Dinand and Simmons, Adrian and Soci, Cornel and Abdalla, Saleh and Abellan, Xavier and Balsamo, Gianpaolo and Bechtold, Peter and Biavati, Gionata and Bidlot, Jean and Bonavita, Massimo and De Chiara, Giovanna and Dahlgren, Per and Dee, Dick and Diamantakis, Michail and Dragani, Rossana and Flemming, Johannes and Forbes, Richard and Fuentes, Manuel and Geer, Alan and Haimberger, Leo and Healy, Sean and Hogan, Robin J. and Hólm, Elías and Janisková, Marta and Keeley, Sarah and Laloyaux, Patrick and Lopez, Philippe and Lupu, Cristina and Radnoti, Gabor and de Rosnay, Patricia and Rozum, Iryna and Vamborg, Freja and Villaume, Sebastien and Thépaut, Jean-Noël},
title = {The {ERA5} global reanalysis},
journal = {Quarterly Journal of the Royal Meteorological Society},
volume = {146},
number = {730},
pages = {1999-2049},
doi = {https://doi.org/10.1002/qj.3803},
year = {2020}
}

@misc{hahner2026,
      title={Representing the Surface Ocean in ECMWF's data-driven forecasting system AIFS},
      author={Sara Hahner and Lorenzo Zampieri and Jean-Raymond Bidlot and Philip Browne and Matthew Chantry and Mariana C. A. Clare and Harrison Cook and Peter Dueben and Rachel Furner and Sarah Keeley and Josh Kousal and Simon Lang and Christian Lessig and Gert Mertes and Kristian Mogensen and Gabriel Moldovan and Charles Pelletier and Florian Pinault and Ana Prieto Nemesio and Baudouin Raoult and Irina Sandu and Mario Santa Cruz and Jakob Schloer and Steffen Tietsche and Hao Zuo},
      year={2026},
      eprint={2604.25559},
      archivePrefix={arXiv},
      url={https://arxiv.org/abs/2604.25559},
}

@misc{lehmann2026,
      title={Can AI Weather Models Predict Beyond Two Weeks? A Quantitative Benchmark and Analysis of Long Rollouts},
      author={Fanny Lehmann and Firat Ozdemir and Yun Cheng and Torsten Hoefler and Sebastian Schemm and Benedikt Soja and Siddhartha Mishra},
      year={2026},
      eprint={2605.30184},
      archivePrefix={arXiv},
      primaryClass={cs.LG},
      url={https://arxiv.org/abs/2605.30184},
}

@misc{kasteleyn2026,
      title={PhysMetrics.Weather: An Evaluation Framework for Physical Consistency in ML Weather Models},
      author={Emma Kasteleyn and Timo Maier and Axel Lauer and Veronika Eyring and Pierre Gentine and Ana Lucic},
      year={2026},
      eprint={2606.10642},
      archivePrefix={arXiv},
      primaryClass={cs.LG},
      url={https://arxiv.org/abs/2606.10642},
}
\end{document}